\documentclass[prb,aps,twocolumn,superscriptaddress,floatfix]{revtex4}
\usepackage{graphicx}
\usepackage{bm}
\usepackage{amssymb}

\def\tit#1#2#3#4#5{{#1}{\bf #2}, #3 (#4)}

\begin{document}

\title{A striped supersolid phase and the search for deconfined quantum criticality 
 in hard-core bosons on the triangular lattice}

\author{R. G. Melko}
\affiliation{Materials Science and Technology Division, Oak Ridge National Laboratory,
Oak Ridge, Tennessee 37831}
\author{A. Del Maestro}
\affiliation{Department of Physics, Harvard University, Cambridge, Massachusetts 02138}
\author{A. A. Burkov} 
\affiliation{Department of Physics, Harvard University, Cambridge, Massachusetts 02138}

\date{\today}

\begin{abstract}

Using large-scale quantum Monte Carlo simulations we study bosons hopping on a triangular lattice with nearest ($V$) and next-nearest ($V^{\prime}$) neighbor repulsive interactions.  In the limit where $V=0$ but $V^{\prime}$ is large, we find an example of an unusual period-three striped supersolid state that is stable at 1/2-filling.  We discuss the relationship of this state to others on the rich ground-state phase diagram, which include a previously-discovered supersolid, a uniform superfluid, as well as several Mott insulating phases.  
We study several superfluid- and supersolid-to-Mott phase transitions, including one proposed by a recent phenomenological dual vortex field theory as a candidate for an exotic deconfined quantum critical point.   We find no examples of unconventional quantum criticality among any of the interesting phase transitions in the model.
\end{abstract}

\maketitle

\section{Introduction}

The question of the effects of frustrated repulsive interactions on the Bose Hubbard Hamiltonian is an intriguing one.  Over the last several years, studies of the simplest hard-core boson Hamiltonian on two-dimensional (2D) frustrated lattices have revealed fascinating non-trivial quantum phases.\cite{moessner1}  The two most exciting recent examples are a stable supersolid phase in triangular lattice bosons at 1/2-filling,\cite{sstri1,sstri2,sstri3} and a valence-bond solid (VBS) phase at 2/3-filling on the kagome lattice.\cite{cabra,kagomeDVT,kagomeVBS} The VBS order corresponds to a short-range resonance of bosons around a lattice bond or plaquette, the supersolid phase to a full delocalization of bosons across the lattice, resulting in off-diagonal long-range order at zero-temperature.

The desire to extend these studies to other models in this general class is immediate.  There is particular interest in the community to find other examples of non-trivial bulk quantum phases, and possibly to relate existing ones to more realistic models of cold atoms trapped in optical lattices.   The inclusion of further-neighbor interactions in the Bose Hubbard Hamiltonian is extremely important for efforts\cite{dipolarBEC} that attempt to use it to model a dilute gas of atoms with dipolar interactions (such as $^{52}$Cr )\cite{Griesmaier} 
in a periodic potential. 
In addition, perturbations of the simplest hard-core boson Hamiltonian offer the opportunity to make connection with continuum quantum field theories studying the phenomenology of exotic quantum phase transitions.  In this work, we enlist the results of a recent dual vortex theory (DVT)\cite{theBurkonian} to help perform a ``guided search'' for examples of exotic (non-Landau-Ginzburg-Wilson (LGW)) deconfined quantum critical phenomena.\cite{deconf1} 

We consider a model of bosons on the triangular lattice with nearest-neighbor ({\it nn}) hopping and both nearest and next-nearest neighbor ({\it nnn}) repulsion,
\begin{eqnarray}
  H_{b} = &-&t \sum_{\langle i,j \rangle} 
    \left( b^{\dag}_i b^{\phantom\dag}_j + b^{\phantom\dag}_i b^{\dag}_j \right)
    + V \sum_{\langle i,j \rangle } n_i n_j  - \mu \sum_{i} n_i \nonumber \\
    &+& V^{\prime} \sum_{\langle \langle i,j \rangle \rangle} n_i n_j ,
    \label{VtHam}
\end{eqnarray}
where $b^\dagger_i$ ($b^{\phantom\dagger}_i$) creates (destroys) a particle on site $i$, $n_i = b^\dagger_i b^{\phantom\dag}_i$, $t>0$ and $V>0$ denote the {\it nn} hopping and repulsion, $V^{\prime}>0$ is a {\it nnn} repulsion, and $\mu$ is the chemical potential.   
In the usual way ($S^z_i = n_i - 1/2, S^+_i = b^{\dag}_i$ etc.), the model can be mapped to the spin-$1/2$ XXZ model on the triangular lattice,
\begin{eqnarray}
  H_{\text{s}} = &-&J_{\perp}\sum_{\langle i,j \rangle} ( S^+_iS^-_j + S^-_iS^+_j ) + 
  J_z\sum_{\langle i,j \rangle }S^z_i S^z_j \nonumber \\
  &+& J_z^{\prime} \sum_{\langle \langle i,j \rangle \rangle} S^z_i S^z_j - h\sum_i S^z,
  \label{XXZeqn}
\end{eqnarray}
where $J_{\perp}$ is the in-plane exchange, $J_z$ and $J_z^{\prime}>0$ denote the diagonal {\it nn} and {\it nnn} exchange, respectively,
and $h$ is an external magnetic field. The model parameters are related by
$J_{\perp}=2t$, $J_z=V$, $J_z^{\prime}=V^{\prime}$, and $h=\mu-3V-3V^{\prime}$.

This paper is organized as follows: In Section \ref{Landau} we review what is known about this class of Hamiltonian, including past work and general theoretical expectations.  In Section \ref{QMC} we discuss recent progress and implementations of quantum Monte Carlo (QMC) algorithms for the $t$-$V$-$V^{\prime}$-$\mu$ model (Eq.~(\ref{VtHam})) taking special note of efforts in alleviating some common ergodicity problems.  QMC results are presented in two sections: first, results at 1/2-filling ($h=0$) for the $t$-$V$-$V^{\prime}$ model, with a focus on the full phase diagram and in particular supersolidity in the {\it nnn} case ($V=0$) (Section \ref{halffilled}).  In this section we also study several superfluid- and supersolid-to-Mott phase transitions with the aim of searching for examples of interesting quantum phase transitions.  Second, results are presented in Section~\ref{sfmottPT} for the $V=0$, $t$-$V^{\prime}$-$\mu$ model away from 1/2-filling.  There we are able to access an example of a superfluid-to-Mott insulator quantum phase transition, which was suggested by a recent dual vortex study\cite{theBurkonian} to be a candidate for exotic non-LGW type behavior.   In our case we find that this phase transition is strongly first-order.

\begin{figure}[floatfix]
{
\includegraphics[width=3in]{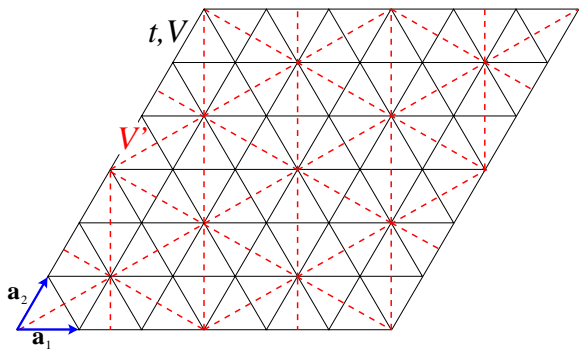}
\caption{(color online) A finite-size $6 \times 6$ ($N=36$) triangular lattice (with periodic boundaries) with one of three {\it nnn}-interacting sublattices illustrated.  Our conventions for primitive vectors are labeled ${\bf a}_1$ and ${\bf a}_2$.
\label{NNNTlatt}}
}
\end{figure}

\section{Past work and general considerations}
\label{Landau}

Various limiting cases of the $t$-$V$-$V^{\prime}$-$\mu$ model are well understood.  
When $t$ dominates, the presence of a uniform superfluid phase has been tacitly established.
The simple {\it nn} Hamiltonian (with $V^{\prime}=0$) has recently been studied by several groups using numerical techniques,\cite{sstri1,sstri2,sstri3,sstri4} where the most interesting feature of the phase diagram is a phase transition out of the uniform superfluid into a supersolid phase (with coexisting superfluid and diagonal long-range order) at $V/t=9$.  This supersolid phase can be viewed as arising from an order-by-disorder mechanism, where an extensive degeneracy of the classical ground states of the interacting part of the Hamiltonian is lifted by quantum fluctuations (hopping).\cite{ClassIsing}
The extensively-degenerate manifold consists of spin configurations that allows for each triangular lattice plaquette to be minimally frustrated (one spin up and two spins down, or vice versa). 
In the supersolid phase, these
configurations obtain diagonal long-range order when perturbed by $t$.  Alternatively, the supersolid state may be viewed as 
resulting from doping the 1/3 or 2/3 (boson density) filled Mott phases that occur at $|\mu |>0$  with additional superfluid bosons.

The purely {\it nnn} model, $V=0$, is expected to share some analogies with the above {\it nn} model.  In the $t=0$ classical limit the three {\it nnn} interacting sublattices are entirely decoupled, leading each {\it nnn} sublattice to realize the frustrated classical groundstate of the equivalent {\it nn} triangular Ising model.   An interesting question that immediately arises is whether the inclusion of a {\it nn} hopping $t$ will promote a similar order-by-disorder effect in the quasi-long-range ($1/\sqrt{r}$) diagonal density-density correlations.  Such a question is non-trivial, since although present in the aforementioned {\it nn} triangular lattice model, a supersolid phase is absent in the analogous kagome lattice model where $1/\sqrt{r}$ correlations do not exist in the frustrated classical manifold.\cite{kagomeDVT,kagomeVBS}  We address this question using QMC simulations in Section \ref{halffilled} of this paper.

\begin{figure}
\includegraphics[width=2.1in]{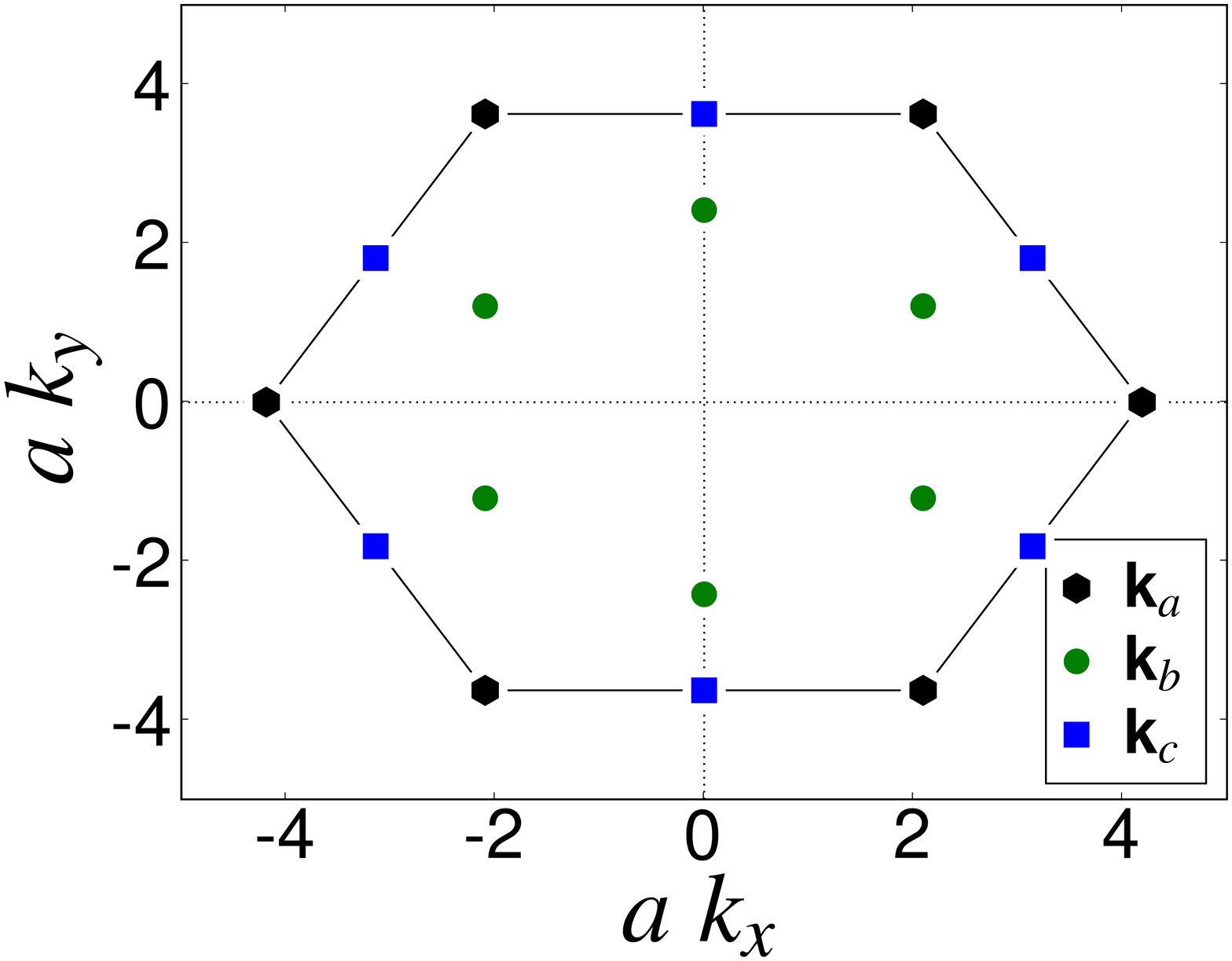}
\caption{(color online)  The three sets of ordering wavevectors for large repulsive interactions discussed throughout the text.
\label{kvecs}}
\end{figure}
To gain insight into the types of order which may develop when large repulsive interactions ($V$ and $V'$) are present, we appeal to a simple calculation of the  ordering wavevectors for the classical {\it nn} and {\it nnn} Ising model.  The ``star'' of ${\bf k}$-points that minimize the Fourier transform of the classical interaction matrix are displayed in Fig.~\ref{kvecs}
and have the following values: for $V^{\prime}/V=0$,
\begin{eqnarray*}
	 {\bf k}_{a_1} &=& (2\pi/3,-2\pi/\sqrt{3}) \nonumber \\ 
	 {\bf k}_{a_2} &=& (4\pi/3,0),
\end{eqnarray*} 	
(and negatives, where the third can be constructed by adding a reciprocal lattice vector) which correspond to the corners of the first Brillouin zone; for $V/V^{\prime}=0$
\begin{eqnarray*}
	 {\bf k}_{b_1} &=& (2\pi/3,-2\pi/3\sqrt{3}) \nonumber \\
	 {\bf k}_{b_2} &=& (0,4\pi/3\sqrt{3}) \nonumber \\
	 {\bf k}_{b_3} &=& (2\pi/3,2\pi/3\sqrt{3});
\end{eqnarray*}
and for $V=V'$:
\begin{eqnarray*}	 
	 {\bf k}_{c_1} &=& (\pi,-\pi/\sqrt{3}) \nonumber \\
	 {\bf k}_{c_2} &=& (0,2\pi/\sqrt{3}) \nonumber \\
	 {\bf k}_{c_3} &=& (\pi,\pi/\sqrt{3}).
\end{eqnarray*}
The ${\bf k}_c$ are bisectors of the boundary of the first Brillouin zone, and ${\bf k}_b = 3/2 {\bf k}_c$.
In the QMC, these ordering wavevectors are employed to construct diagonal order parameters using the density structure factor, as discussed in the next section.  

In the presence of a moderate particle-hole symmetry-breaking chemical potential (magnetic field), the classical {\it nn} model on the triangular lattice realizes a $\sqrt{3} \times \sqrt{3}$ (boson density 2/3 or 1/3) pattern with true long-range ordered correlations at wavevectors ${\bf k}_a$.\cite{Murthy97} The same can be expected for the three {\it nnn} sublattices if not coupled by a hopping ($t=0$).  If one combines these three non-interacting {\it nnn} sublattices to form the original lattice illustrated in Fig.~\ref{NNNTlatt}, two possible global patterns result, with ordering wavevectors  ${\bf k}_b$.  These are the ``striped'' and ``bubble'' phases (illustrated in Fig.~\ref{stripebub}), studied previously using numerical techniques by Metcalf.\cite{Metcalf}  Metcalf demonstrated that both phases are stable, energetically equivalent ground-state configurations of the classical {\it nnn} model in the presence of a magnetic field. We therefore can ask whether either of these ordered (or insulating) patterns survive in the system upon inclusion of the {\it nn} hopping $t$.  If so, then the possibility of accessing the associated superfluid-insulating quantum phase transition (say upon tuning $t$) is of exceptional interest.  This phase transition was studied by Burkov and Balents\cite{theBurkonian} using a phenomenological field theory of vortices interacting on the dual (honeycomb) lattice,  
 with the goal of understanding universal phenomena in the vicinity of various superfluid-Mott quantum critical points in microscopic models with the same spatial symmetries.  The simplest DVT analysis for the case of 1/3 (or 2/3) boson filling contains three unique Mott insulating states.  Two of these Mott phases are the striped and bubble states, and are conjectured to be connected by a deconfined quantum critical point (DQCP) to the neighboring superfluid state.  A DQCP is 
the recently proposed bon ton of unconventional 
quantum criticality, not based on the standard LGW concept of the order parameter.\cite{deconf1}  Rather, such DQCPs are described in terms of {\it emergent} degrees of freedom not related to the microscopic symmetries of the adjoining phases, and emergent topological order.  However, since no clear examples of a DQCP have been identified in microscopic models or real materials, the discovery of the relevant superfluid-insulating quantum phase transition in our model would offer the unique opportunity to search for this exotic phenomenon guided by predictions of the same phenomenological field theory from which it was originally proposed.\cite{DVTmottK}  In Section~\ref{sfmottPT} we report QMC evidence for the existence of the striped and bubble phases in the $t-V^{\prime}-\mu$ model and study in detail the relevant superfluid-insulating phase transition.

\section{Notes on QMC Techniques}
\label{QMC}

In an effort to elucidate the phase diagram of the model and study the physical questions mentioned above, we employ extensive numerical simulations in this work, using the Stochastic Series Expansion (SSE) quantum Monte Carlo method pioneered by Sandvik.\cite{SSEanders1,SSEanders2}  The triangular lattice with next-nearest neighbor interactions is illustrated in Fig.~\ref{NNNTlatt}.  Here, the {\it nnn} diagonal interactions form three decoupled triangular sublattices, which are coupled by the {\it nn} hopping term.
On frustrated lattices such as this, the QMC is free from the sign problem only for positive (ferromagnetic) $t$.  The sign of $V$ or $V^{\prime}$ is not relevant in this regard, however as observed in studies on other frustrated lattices, simulation efficiency is severely affected by large repulsive interactions.  Such problems arise due to the difficulty of tunneling between the classically degenerate manifold of minimally-frustrated states; in classical simulations of such systems, non-local cluster or loop moves are often employed in a Monte Carlo.  However for this type of finite-temperature QMC, the analogous cluster moves are highly non-trivial to implement.

One significant algorithmic advance, which combats the tendency of the SSE directed loop algorithm to freeze (for large $V$ or $V^{\prime}$) is to employ six-leg vertices representing triangular lattice plaquettes instead of 4-leg lattice bonds in the formulation of the directed loop algorithm.\cite{plaquette,SergeiPlaq,RGMnnV}  In the traditional SSE formulation,\cite{SSEanders1,SSEanders2} the Hamiltonian is decomposed into a list of two-site bond operators, which are sampled using diagonal and off-diagonal updates in a Monte Carlo procedure.  The off-diagonal update typically employed is an operator loop algorithm called the ``directed loop'' update: closed loops are constructed in the $d+1$ simulation cell according to transition weights determined by the partition function.  After being built, bosons on these closed loops are added or removed with probability $1$.

A difficulty in this traditional bond algorithm occurs for certain parameters, particularly in the large $V$ or $V^{\prime}$ regime of the phase diagram (the frustrated near-degenerate manifold with small hopping).  As described in Ref.~\onlinecite{RGMnnV}, when the loop under construction encounters a frustrated bond with some diagonal operator, the non-zero transition weight associated with 
adding or removing a single boson (resulting in the sampling of another diagonal operator) is suppressed due to the repulsive interactions.

This difficulty can be overcome by a removal of the ``local'' energy barrier directly associated with this suppression of transition weight.  To do so, one may consider a  decomposition of the Hamiltonian into a list of three-site triangular plaquettes.  The plaquette algorithm can be expected to increase the efficiency of the directed-loop updates; movement between different minimally-frustrated plaquettes are no longer suppressed because both the initial and final configurations of the plaquette have the same (or similar) transfer weights.  The result is a removal of an energy barrier directly associated with traversing local configurations in the near-degenerate frustrated manifold for states located in the large
$V$ or $V^{\prime}$ regime of the phase diagram, and a significant improvement in efficiency for the algorithm.\cite{RGMnnV}

We employ the triangular plaquette version of the SSE algorithm for both {\it nn} and {\it nnn} Hamiltonian terms.  As discussed below, we nonetheless observe ``freezing'' of the QMC dynamics for $V^{\prime}/t > 5$, due to the persistence of ``global'' energy barriers associated with the near-degeneracy of the entire minimally-frustrated manifold and not alleviated by the above algorithmic advances.  This freezing problem could be further reduced for larger $V$ or $V^{\prime}$ by employing more advanced simulations techniques, such as parallel tempering;\cite{RGMnnV} however, we find that the essential physics is captured for the parameter range accessible in this simulation scheme, so we leave such improvements for future studies.

The standard estimator for the superfluid density of the model is the stiffness,\cite{stiffy}
defined via the free energy response to a twist $\phi$ in the boundary conditions and measured with winding number fluctuations\cite{windingnumber} in each of the lattice translation vector directions, 
\begin{equation}
\rho_s^{{\bf a}} = \frac{\langle W_{{\bf a}}^2 \rangle}{\beta},
\label{RHOs}
\end{equation}
where $\beta = t/T$ is the inverse temperature.
Typically the stiffness is averaged over both ${\bf a}_1$ and ${\bf a}_2$ directions (see Fig.~\ref{NNNTlatt}) so that we define $\rho_s=(\rho_s^{{\bf a}_1}+\rho_s^{{\bf a}_2})/2$,
unless measured in a phase which breaks rotational symmetry (we discuss this more in the next section).

The presence of a non-zero spin stiffness in the thermodynamic limit indicates off-diagonal long-range order, manifest as either a uniform superfluid or supersolid phase.  The standard indicator for the presence of diagonal long-range order is the density structure factor, 
\begin{equation}
S({\bf k}) = \frac{1}{N} \sum_{i,j}
{\rm e}^{i\,{\bf k}\cdot({\bf r}_i - {\bf r}_j)}
\big(\langle n_i n_j \rangle - \langle n_i \rangle \langle n_j \rangle \big),
\label{Szstr}
\end{equation}
easily measured in the $n_i$ basis, which is diagonal in our SSE implementation.  The order parameter for diagonal correlations is constructed from the structure factor by noting the non-zero ordering wavevectors ${\bf k}_{\alpha_i}$ through the definition
\begin{equation}
\left\langle m_{\alpha_i}^2 \right\rangle = \frac{S({\bf k}_{\alpha_i})}{N},
\label{OP}
\end{equation}
where $N$ is the finite number of lattice points. Any stable phase having non-zero values for both $\rho_s$ and $\langle m_{\alpha_i}^2\rangle$ in the thermodynamic limit can be classified as a supersolid,
while we refer to the phase with only $\langle m_{\alpha_i}^2\rangle \neq 0$ as a Mott insulator.

\section{half-filling}
\label{halffilled}

\subsection{Next-nearest neighbor striped supersolid}

\begin{figure}[floatfix]
{
\includegraphics[width=3.2in]{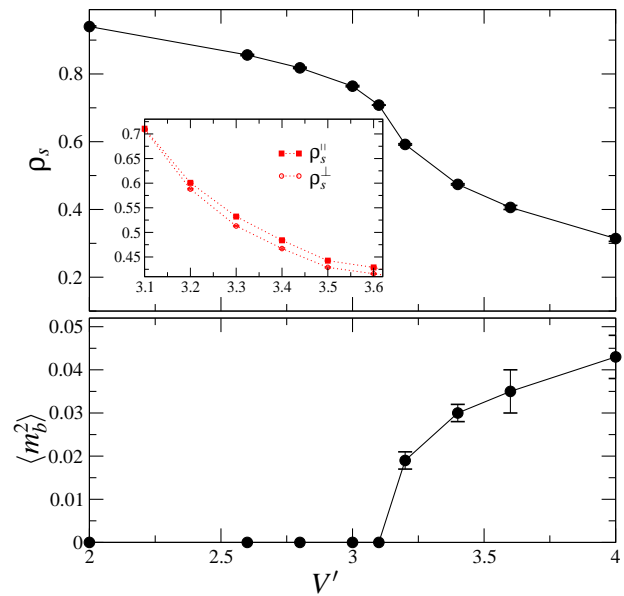}
\caption{(color online) Superfluid-supersolid phase transition.  The top panel illustrates the superfluid density ($\rho_s=(\rho_s^{{\bf a}_1}+\rho_s^{{\bf a}_2})/2$), and the bottom the solid order parameter, as a function of the {\it nnn} repulsion.   Data for both main plots has been extrapolated to the thermodynamic limit.  
The inset shows a detail of the superfluid density in the supersolid phase.  The values $\rho_s^{\shortparallel}$ and $\rho_s^{\perp}$ are the superfluid density parallel and perpendicular to the stripe direction, discussed in the text.
\label{SFSS}}
}
\end{figure}
\begin{figure}[floatfix]
{
\includegraphics[width=2.7in]{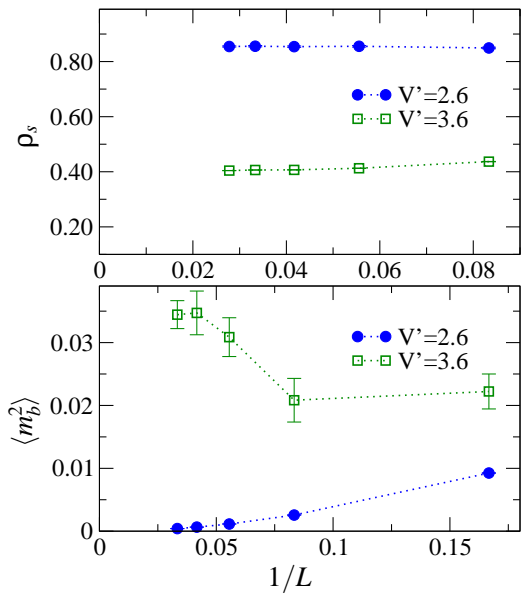}
\caption{(color online)  Examples of finite-size scaling for quantities that appear in the main plot
of Fig.~\ref{SFSS} for two values of $V^{\prime}$.
\label{SFFSS}}
}
\end{figure}

We begin by studying the $V=0$ Hamiltonian at 1/2-filling, or $\mu = 3V'$ (corresponding to zero magnetic field in the XXZ model).   As described in Section~\ref{Landau}, in the classical limit ($t=0$) the existence of the nnn $V^{\prime}$ interaction alone produces three decoupled triangular lattices, each of which realizes the usual ground state described by the set of wavevectors ${\bf k}_{b_i}$.  It is most easily described as a degenerate manifold of minimally-frustrated triangular plaquettes, each with 
only one or two bosons.
It is expected that the addition of a {\it nn} hopping ($t$) will break the classical degeneracy, although whether this perturbation produces an ordering effect on the diagonal components of the bosons is not obvious.  

We can address this physical question with QMC simulations, however, pragmatically, the approach of perturbing the classical $V^{\prime}$ manifold with $t$ is difficult due to the aforementioned severe algorithmic freezing effects in this regime.  We thus employ the usual complementary approach, by applying a strengthening diagonal $V^{\prime}$ interaction to the dominant hopping regime (we use a fixed $t=0.5$ corresponding to $J_{\perp} =1$ throughout this paper).  This allows us to study the effects of increasing frustration while simultaneously evaluating the simulation performance in a systematic way.  Similarly, ground-state estimators are obtained by the usual method of cooling the simulation cell (by lowering $T$ systematically) until convergence is reached to within error bars.  Since algorithmic freezing strongly affects error bars, a delicate balance must be struck in order to find a temperature that faithfully represents the ground-state physics, while at the same time retaining ergodicity on reasonable simulation time-scales.

\begin{figure}[floatfix]
{
\includegraphics[width=3.2in]{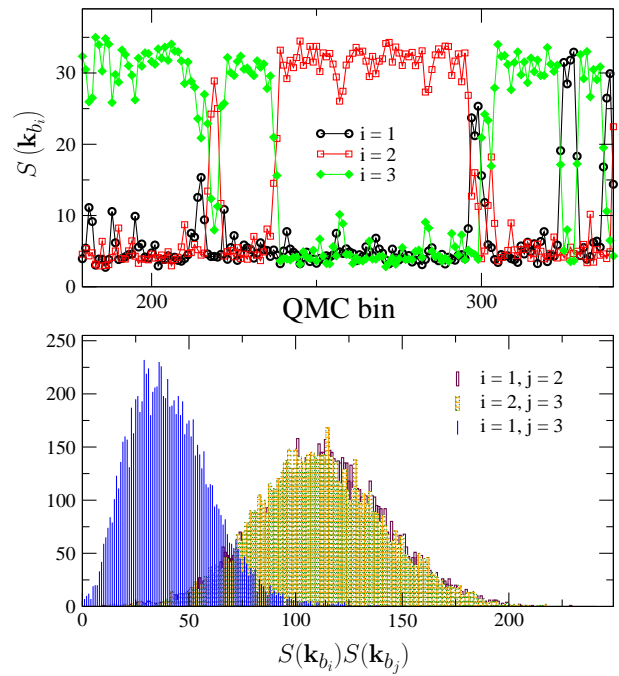}
\caption{(color online) 
The three {\it nnn} striped diagonal order parameters ($S({\bf k}_{b_{i}})$) plotted as a function of Monte Carlo time (top) for a $30 \times 30$ lattice.  Each data point is a bin average of $10^4$ QMC steps.  A histogram of the three possible product combinations (bottom) for single bin of $10^4$ QMC steps indicates the presence of striped correlations, as discussed in the text.  The stiffness values calculated for this same simulation were $\rho_s^{{\bf a}_1}=0.398(8)$ and $\rho_s^{{\bf a}_2}=0.384(8)$.
\label{histo}}
}
\end{figure}

Results of the simulations are illustrated in Fig.~\ref{SFSS}.  As one increases the frustrating
$V^{\prime}$, we observe a phase transition out of the usual ``uniform'' superfluid phase at dominant
$t$.  This phase transition is indicated by a drop in the $\rho_s$ curve in Fig.~\ref{SFSS} at
$V^{\prime} \approx 3$, while $\rho_s$ remains finite (in the thermodynamic limit) on both sides of the transition.  
Note that $\rho_s$ is measured in units of $t$, which has been set to 0.5 throughout this work.
For  $V^{\prime} > 3$, long-range order develops in the diagonal density correlations, as indicated by very sharp Bragg peaks in the structure factor $S({\bf k})$ at ordering wavevectors ${\bf k}_{b_i}$ which correspond to the expectations from arguments involving the symmetry of the interactions in the {\it nnn} Ising model (Section~\ref{Landau}). The diagonal long range order is classified using the set of order parameters $\langle m_{b_i}^2\rangle$ (Eq.~(\ref{OP})).  The large $V^{\prime}$ phase exhibits concurrent diagonal long range order (as indicated by the survival of the order parameter $\langle m_{b}^2\rangle$ in the thermodynamic limit) and off-diagonal long range order (indicated by a non-zero superfluid density) signaling that the entered phase is a thermodynamically stable and homogeneous supersolid at 1/2-filling.  
The survival of $\rho_s$ and $\langle m_b^2 \rangle$ are rigorously confirmed using finite size scaling, illustrated in in Fig.~\ref{SFFSS}.   We observe that $\rho_s$ quickly converges to its thermodynamic limit value, while $\langle m_b^2 \rangle$ requires the simulation of relatively large lattices in the supersolid phase -- this apparent non-monotonicity in the scaling is discussed further below.  In the uniform superfluid phase, $\langle m_b^2 \rangle$ extrapolates to zero as expected for a phase without diagonal long range order.  Deeper in the supersolid phase, for $V^{\prime}>5$, we observe that simulation efficiency is quickly destroyed, manifest as observations of non-equilibrium freezing in the measured observables.

The ordering wavevectors at ${\bf k}_{b_i}$ are consistent with a striped or bubble phase as observed
in the classical {\it nnn} Ising model (see Section~\ref{Landau}), and could be expected in a canonical simulation fixed at 1/3 or 2/3-filling.\cite{Metcalf}  Our simulations are however grand-canonical and the observed supersolid is firmly at 1/2-filling, as indicated by studying histograms of the particle density.  
The supersolid is homogeneous and shows no obvious evidence of phase separation.

In a phase such as a supersolid, with hopping bosons delocalized throughout the lattice, one cannot distinguish the particular diagonal sublattice ordering in QMC simulations by looking directly at density ``snapshots''. Any obvious hallmarks of ordering patterns tend to be washed out by the hopping.  We can however learn more about the sublattice ordering by studying the symmetry of the ordering wavevectors.  Specifically, in the thermodynamic limit, the striped phase (see also Fig.~\ref{stripebub}) is characterized by the existence of {\it only one} nonzero value for $\langle m_{b_i}^2\rangle$, whereas the bubble phase has all {\it three} components nonzero and equal in magnitude.  Monte Carlo data for the three independent $S({\bf k}_b)$ is illustrated in Fig.~\ref{histo} for a large system size in the supersolid phase.  It is clear that the three order parameters ``trade off'' between a large and small value.  If one remembers that $S({\bf k}_b(L))$ retains some non-zero magnitude even in a disordered phase due to finite size effects, it is apparent that the behavior in Fig.~\ref{histo} suggests a stripe-like symmetry.  Further evidence can be obtained by looking at histograms of the products of the order parameters: $S({\bf k}_{b_1}) S({\bf k}_{b_2})$, $S({\bf k}_{b_2}) S({\bf k}_{b_3})$, and $S({\bf k}_{b_1}) S({\bf k}_{b_3})$.  In a finite-size remnant of the striped phase, one expects one of these products to be small and the other two large.  The bottom plot in Fig.~\ref{histo} is clearly consistent with this expectation.
With this in mind, we note that one must be careful to include only the largest $S({\bf k}_b)$ estimator in the construction of the striped order parameter (for example in Fig.~\ref{SFSS}).  We observe that below a system size of $12 \times 12$, no single element of $S({\bf k}_b)$ seems to be preferred. 
The non-monotonicity in the finite-size scaling of $\langle m_{b}^2\rangle$ (for $V^{\prime}=3.6$) in Fig.~\ref{SFFSS} apparently occurs at the point where the finite simulation cell becomes large enough to see the length scale associated with stripe development.  This interesting feature should be investigated further.

The above evidence points to the naive picture of the supersolid phase as a doped version of the 1/3 or 2/3 period-three striped Mott phase.  However some of the behavior of the supersolid is unusual in this regard.  In particular, as alluded to above (see Fig.~\ref{histo}), $\rho_s$ does not show strong anisotropy or correlation with stripe ``direction'' (i.e.~the largest value of $S({\bf k}_{b_1})$, $S({\bf k}_{b_2})$ or $S({\bf k}_{b_3})$)
 such as one might expect for a system with broken rotational symmetry.  
To study this further, we observe $\rho_s$ along both primitive lattice directions during the QMC run.   In a strongly developed striped phase with large $S({\bf k}_{b_1})$, one might expect the parallel-running superfluid density, $\rho_s^{{\bf a}_2}$ to be largest, while in a $S({\bf k}_{b_2})$ striped phase, $\rho_s^{{\bf a}_1}$ might dominate.  We ran simulations that, for each QMC configuration sampled, binned $\rho_s^{{\bf a}}$ into a new estimator $\rho_s^{\shortparallel}$ or $\rho_s^{\perp}$.  If stripes running parallel to ${\bf a}_1$ were observed (i.e.~$S({\bf k}_{b_2})$ is largest), $\rho_s^{\shortparallel}=\rho_s^{{\bf a}_1}$ and $\rho_s^{\perp}=\rho_s^{{\bf a}_2}$; vice versa for stripes running parallel to ${\bf a}_2$.  If stripes were observed to run parallel to the third symmetry direction, i.e.~$S({\bf k}_{b_3})$ is largest, no data for $\rho_s^{\shortparallel}$ or $\rho_s^{\perp}$ was binned, since one might expect the superfluid density to be isotropic in this case.

We explore the possibility for $\rho_s$ anisotropy in the striped supersolid phase.
Data for a region of the supersolid phase adjacent to the transition is illustrated in the inset to Fig.~\ref{SFSS}, taken for a $30 \times 30$ system at $\beta=12.5$
This measurement indicates that statistically significant anisotropies are present,
with deviations of a few percent, for example at $V^{\prime}=3.4$, where $\rho_s^{\shortparallel} = 0.484(2)$ and $\rho_s^{\perp}=0.467(3)$.
However, it is clear that large anisotropies like those 
reported in different striped supersolid states in Bose Hubbard models on the square 
lattice\cite{OtherStripe} are not evident in our {\it nnn} triangular striped supersolid.
This suggests that the usual interpretation of the supersolid as a doped Mott insulator may not be applicable here.
We revisit this discussion after looking at the period-three striped Mott phase in Section~\ref{sfmottPT}.

\subsection{$V$-$V'$ phase diagram}

\begin{figure}[floatfix]
{
\includegraphics[width=3.2in]{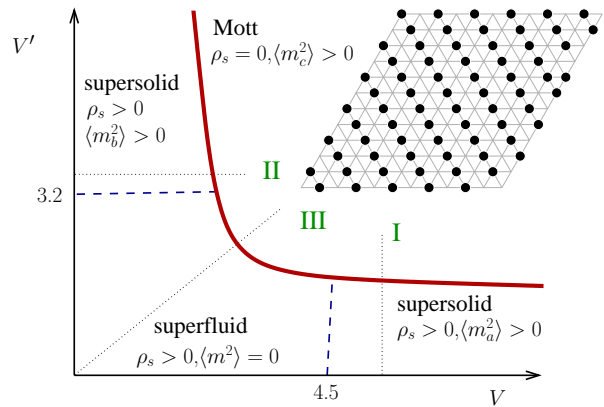}
\caption{(color online) 
Schematic ground-state phase diagram for the zero-field $t$-$V$-$V^{\prime}$ model.  Blue (dashed) lines represent transitions between the superfluid and supersolid states.  The red (thick) line is the phase boundary between these phases and the period-two Mott insulator, which is strongly first order.  Dotted lines labeled I, II and III illustrate the position of data sweeps in the next three figures.  The inset shows one of the possible period-two striped Mott phases, found by plotting the boson density for an instantaneous QMC configuration with $V = 5.0$ and $V' = 0.8$.
\label{VVpPhaseD}}
}
\end{figure}

Now that we have firmly established the presence of a {\it nnn} striped supersolid phase at large $V^{\prime}$ that exists complementary to the previously discovered {\it nn} supersolid, it is natural to extend our exploration of the phase diagram by perturbing into the $V$, $V^{\prime}>0$ region.  An interesting Mott state is uncovered, which exists at 1/2-filling for  moderate to large $V$, $V^{\prime}>0$.  The phase is a period-two striped state (see Fig.~\ref{VVpPhaseD}), with boson density modulated at one of the ordering wavevectors ${\bf k}_c$ = $(\pi,-\pi/\sqrt{3})$, $(0,2\pi/\sqrt(3)$ or $(\pi,\pi/\sqrt{3})$.  These wavevectors are the same as those expected from the minimization of the ground state energy for the classical ($t=0$) limit of the model in this parameter regime, as discussed in Section~\ref{Landau}.

The phase boundaries between this Mott state and the superfluid and supersolid phases in Fig.~\ref{VVpPhaseD} are interesting examples of quantum phase transitions that we can characterize using the QMC simulation.   Data along three cuts in the phase diagram in Fig.~\ref{VVpPhaseD} are discussed in detail below.

\begin{figure}[]
{
\includegraphics[width=3in]{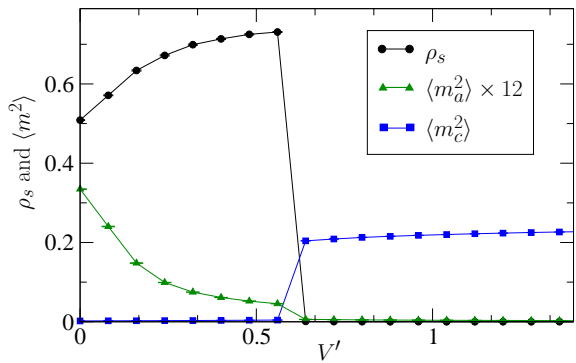}
\caption{(color online)
Superfluid density, {\it nn} supersolid order parameter, and period-two Mott order parameter along cut I in the phase diagram.  Data was taken at $V=5$ for a $L= 12$ lattice at $\beta=12.5$.
\label{Jz5SFMott}}
}
\end{figure}

First, we examine cut I by starting the simulation in the {\it nn} supersolid phase (at a value of $V=5$) and ``perturbing'' away from the $x$-axis by increasing $V^{\prime}$ (Fig.~\ref{Jz5SFMott}).  It is evident upon increasing $V^{\prime}$ that the magnitude of the {\it nn} diagonal order parameter ($\langle m^2_a \rangle$) decreases until terminating in a phase transition, upon which $\rho_s$ falls abruptly to zero and the period-two Mott order parameter ($\langle m^2_c \rangle$) jumps almost to its maximum value.  The sharp discontinuity at the phase boundary is fairly typical of a strong first-order transition.

\begin{figure}[]
{
\includegraphics[width=3in]{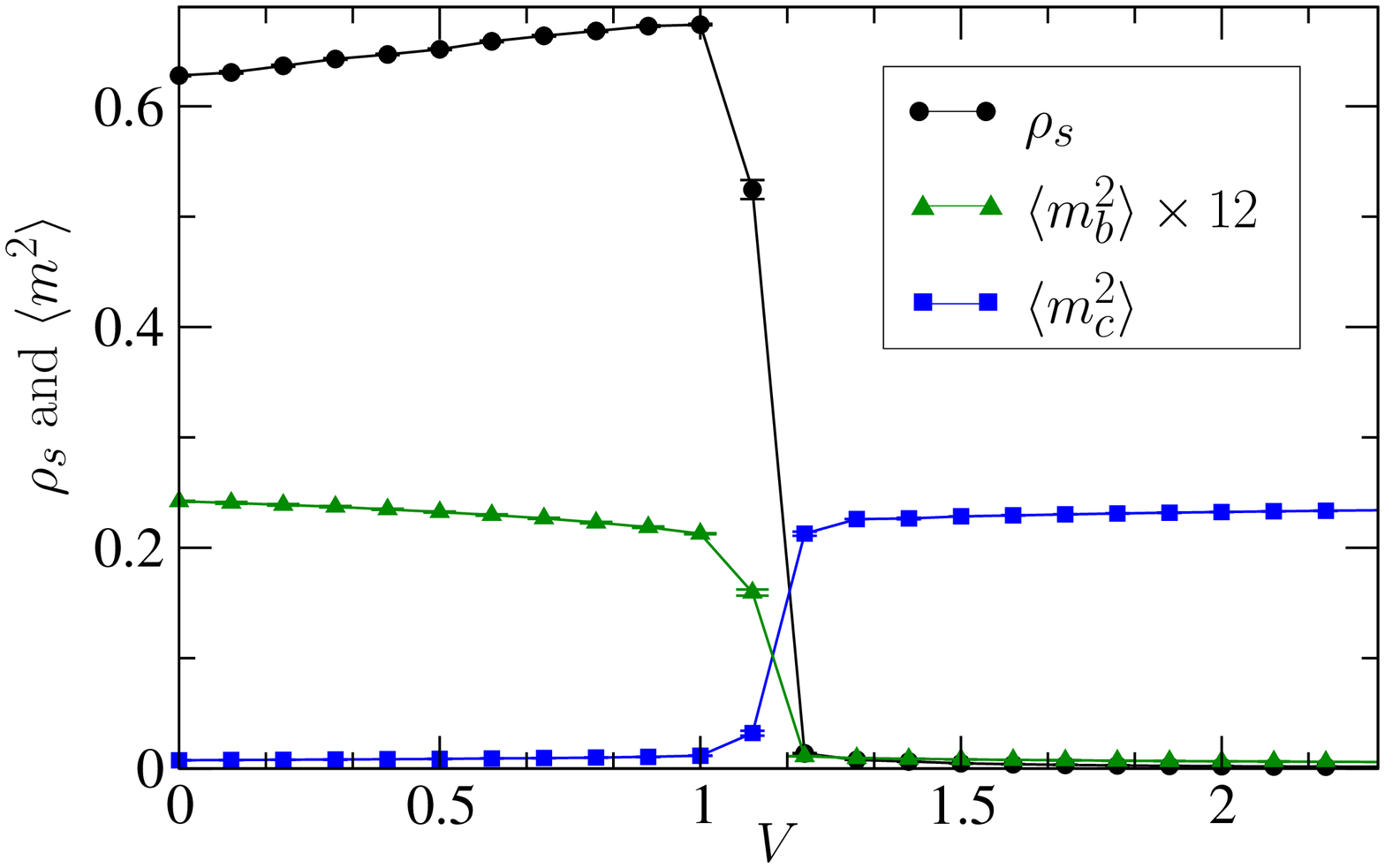}
\caption{(color online) 
Superfluid density and period-two Mott order parameter along cut II in the phase diagram.  Data was taken for a $L= 6$ lattice at $V^{\prime}=3.4$ and $\beta=10$.
\label{Jp3p4}}
}
\end{figure}

A qualitatively similar behavior occurs along cut II as we perturb the {\it nnn} striped supersolid at $V^{\prime}=3.4$ by the interaction $V$.   In this case, a much smaller lattice was employed due to the observation of severe algorithmic freezing in the simulation for moderate to large values of $V$ and $V^{\prime}$.  Nonetheless, the abrupt jumps in the observables in Fig.~\ref{Jp3p4} are suggestive of a finite-size remnant of a strong first-order phase transition, even for a system size as small as $6 \times 6$.

\begin{figure}[]
{
\includegraphics[width=3in]{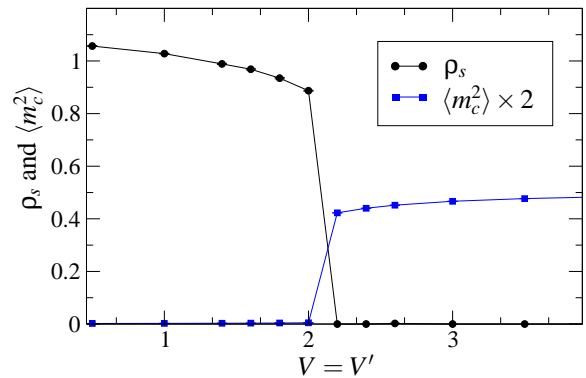}
\caption{(color online)
Superfluid density and period-two Mott order parameter along cut III in the phase diagram.  Data was taken for a $L= 18$ lattice at $\beta=12.5$.
\label{JzJz}}
}
\end{figure}

Finally, we examine the remaining region of the phase boundary by beginning the simulation in the uniform superfluid phase ($V=V^{\prime}=0$) and perturbing by increasing $V=V^{\prime}$.  We again encounter an apparent direct first order phase transition between the superfluid and period-two Mott state, with no other obvious intervening phases occurring (Fig.~\ref{JzJz}).  

The superfluid- and supersolid-to-Mott phase boundaries in Fig.~\ref{VVpPhaseD} appear to be simple, strongly first-order phase transitions.  This result is perhaps expected in a typical LGW paradigm since very different symmetries are broken on each side of the phase boundary.  In contrast, we turn now to an example of a superfluid-to-Mott quantum phase transition where the theoretical possibility exists for exotic non-LGW behavior.

\section{Away from 1/2-filling, $\mu>3$ and $V=0$}
\label{sfmottPT}

\begin{figure}[floatfix]
{
\includegraphics*[width=3.1in]{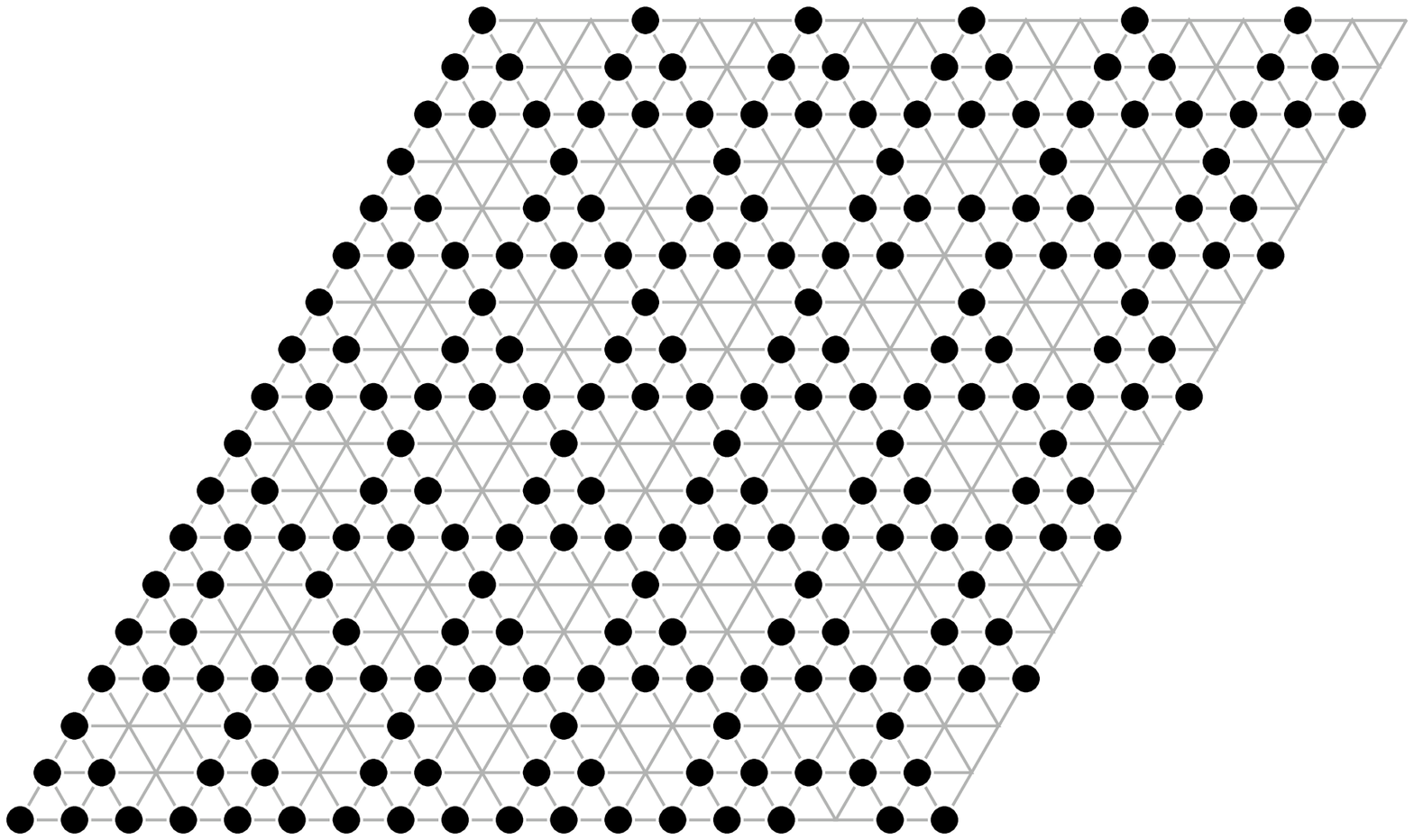}
\includegraphics*[width=3.0in]{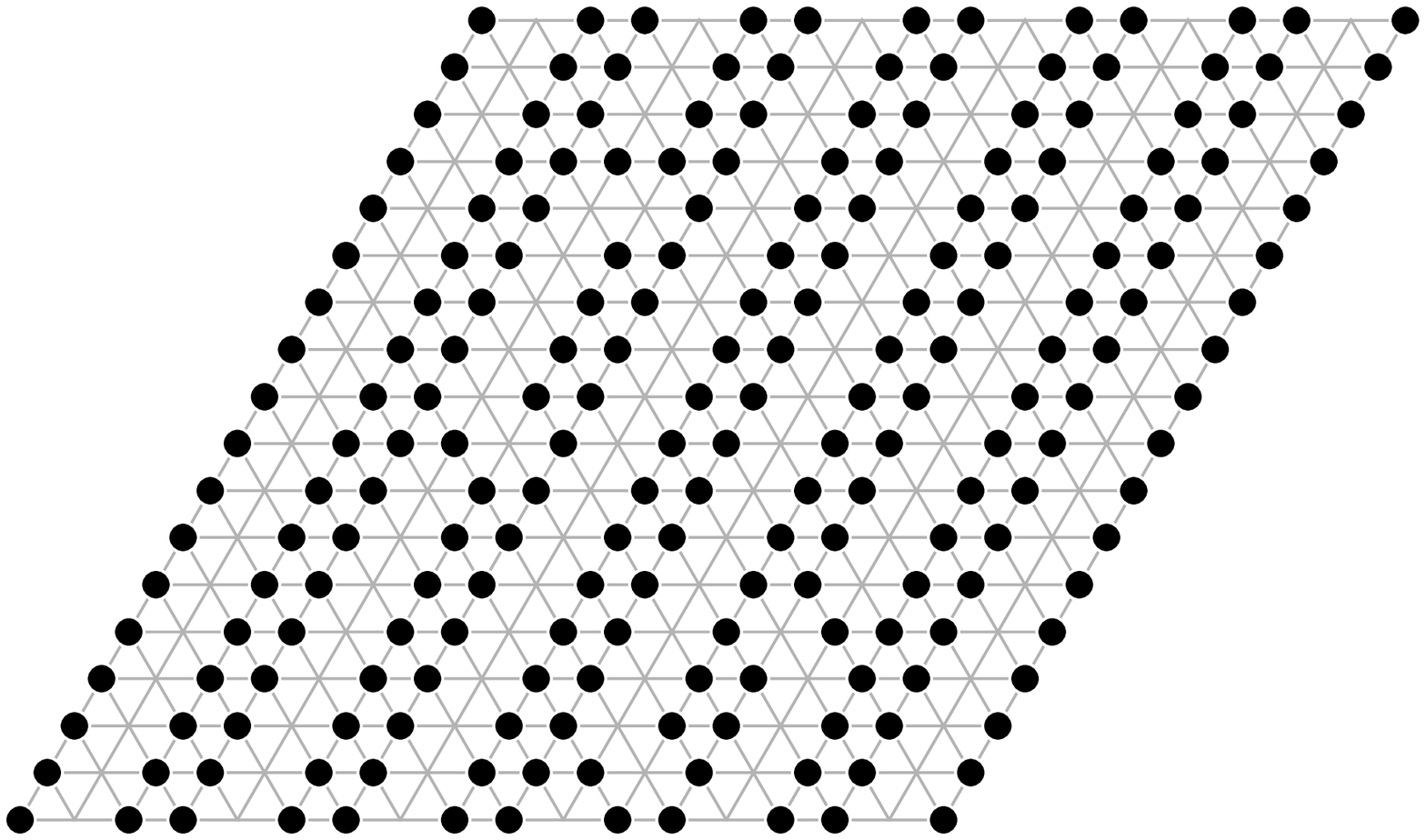}
\caption{(color online) 
QMC ``snapshots'' of instantaneous boson density (spin) arrangements 
stabilized on a $18 \times 18$ simulation cell for $\beta=12.5$, $V'=2.75$ and $\mu=-4.5$. The defects seen in these pictures are not static.  Configurations shown have density 2/3, and are referred to in the text as the ``bubble'' (top) phase and ``striped'' (bottom) phase.  
\label{stripebub}}
}
\end{figure}

Consider the model Hamiltonian Eq.~(\ref{VtHam}) with the {\it nn} repulsion set to zero ($V=0$), and $V^{\prime}$ large enough to be inside the {\it nnn} supersolid phase.
The application of a chemical potential to tune the supersolid phase away from 1/2-filling reveals interesting features.  The phase diagram is observed to have the usual ``lobe'' structure, characteristic of models with frustrated boson repulsion.  Explicit illustrations of this phase diagram can be found in Refs.~\onlinecite{sstri1}, \onlinecite{kagomeVBS} and \onlinecite{Murthy97}.  Of particular interest are the two lobes of Mott insulator at fillings 1/3 and 2/3, which in the present case are separated by the striped supersolid state.  These Mott insulating regions are observed to have the same ordering wavevectors as the supersolid (${\bf k}_b$).  Further, the bosons in these Mott phases are strongly localized, allowing one to observe the sublattice density directly from ``snapshots'' of the diagonal QMC simulation basis, something that is not possible in the supersolid phase.

Interestingly, in the simulations the Mott insulating phases appear able to manifest both expected (period-three striped or bubble) sublattice ordering patterns, depending on equilibration details, and what starting configurations were used in the QMC.  Such behavior occurs due to an algorithmic non-ergodicity that may cause ``sticking'', i.e.~a suppression of free sampling in these two low-lying states.
Examples of the states are illustrated in Fig.~\ref{stripebub}, where two independent simulations were run for parameters inside of the 2/3 Mott lobe starting from different random basis configurations.  One simulations settles firmly in the striped state, the other in the bubble state.  After fairly long run-times ($10^8$ QMC steps), we do not observe any ``mixing'' between states, i.e. one simulation remains in the striped state and the other remains in the bubble state.  Although boson hopping processes are not entirely suppressed (indeed the ``defects'' illustrated in Fig.~\ref{stripebub} are not fixed), large energy barriers exist that prevent these two examples of sublattice ordering from mixing.  The preference of the model towards either state may be delicate, or vary with position on the phase diagram. 
Nonetheless we observe that the average energy of the striped state $E=-2.72306(4)$ (measured in units of $t=0.5$) is slightly lower than the energy of the bubble state $E=-2.72280(4)$ for the specific parameters discussed in Fig.~\ref{stripebub}.  It is interesting to note that a preference of the system towards the 2/3 striped state is 
consistent with the picture of this phase being closely connected to the adjacent {\it nnn} supersolid phase, i.e.~upon crossing the Mott to supersolid phase boundary, the broken symmetry of the diagonal order parameter stays the same.

\begin{figure}[floatfix]
{
\includegraphics[width=3in]{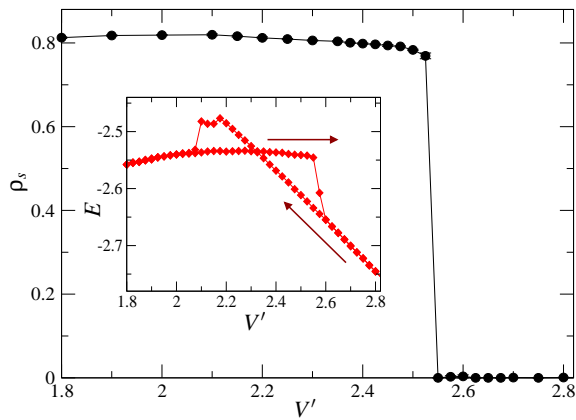}
\caption{(color online) The superfluid-Mott transition at 2/3 filling, for a $L=18$ system.  The simulation parameters were $\beta=12.5$, and $h=3.750$ ($\mu=h-3V'$).  This value of $h$ was found to very closely match the filling of the superfluid right at the transition to 2/3.
The main plot illustrates the abrupt discontinuity in $\rho_s$ characteristic of a first-order transition.  The inset shows hysteresis typical of a first-order energy level crossing.
\label{SFmott}}
}
\end{figure}

As discussed in Section~\ref{Landau}, a recent DVT\cite{theBurkonian} predicts that two specific 2/3 Mott states are connected to a neighboring superfluid (at the same filling) by an exotic DQCP.  These two Mott states are the striped and bubble states, both of which are found in our QMC simulations as described above.  We can easily access the superfluid-Mott phase boundary by varying model parameters ($h$ or $V^{\prime}$) in the QMC.
Thus, with our simulation we are in the unique position of possibly making a direct connection between the behavior of a simple microscopic model and an exotic phenomenon contained in a continuum DVT.

The first step in the simulations is to study the order of the phase transition.  The ``conventional'' situation in a case like this where the phases on either side of the phase transition have unrelated broken symmetries is a first-order transition; a continuous quantum critical point would suggest a breakdown of the conventional Landau picture and possible relevance to a more exotic theory like deconfined quantum criticality.\cite{deconf1}

Using the QMC simulation, we mapped out the phase diagram roughly near the ``tip'' of the Mott lobe, in attempt to determine a set of parameters which closely matches the filling of the neighboring superfluid phase to 2/3 (which would be required to make a comparison to the predictions of the fixed-filling DVT).  We then performed more detailed simulations, keeping the chemical potential fixed and sweeping the parameter $V^{\prime}$.  The results are illustrated in Fig.~\ref{SFmott}.  These simulations, even though carried out on a relative small lattice ($18 \times 18$), display a discontinuity in the value of $\rho_s$ and hysteresis effects in other
observables -- behavior that is prototypical of a first-order phase transition.  In fact,
the superfluid-Mott phase transition appears ``strongly'' first-order, e.g. the discontinuity in $\rho_s$ is clear and obvious.  This result precludes an interpretation of the transition as a quantum critical point that should share universal properties with a DQCP, and instead suggests the much more conventional LGW paradigm as the correct descriptor of the transition.

\section{Discussion}

In this paper we have presented QMC simulation results for the ground state phase diagram of the Bose Hubbard model with {\it nn} hopping and {\it nnn} repulsive interactions.  At 1/2-filling, the phase diagram contains a superfluid phase (for dominant hopping ($t$)), two supersolid phases (for dominant $V$ and $V^{\prime}$ respectively), and a period-two striped Mott insulating state for large $V$ and $V^{\prime}$.  The superfluid- and supersolid-to-Mott phase boundaries are observed to be strongly first-order.

The large-$V^{\prime}$ supersolid phase is characterized by the homogeneous coexistence of diagonal (solid) and off-diagonal (superfluid) long range order in the system.  Density correlations in this phase are observed to have the symmetry of a period-three striped state, similar to that observed in an adjacent 1/3-filled striped Mott insulator,
although the density of the supersolid is measured to be firmly at 1/2-filling (for $h=0$).  This fact leads to the immediate interpretation of the supersolid state as a 1/3 Mott insulator which is doped with interstitial bosons that become delocalized, forming the ``superfluid'' component.  This picture is clearly an over-simplification, however, since as a result one would expect the boson superflow to occur in one-dimensional channels between the strongly-localized density stripes.  
Indeed the resulting anisotropy in the superfluid density is measured, although it is extremely small ($\approx 5\%$), especially when compared to other striped supersolid states observed on square-lattice Bose Hubbard models doped away from 1/2-filling.\cite{OtherStripe}  This issue might be answered by looking at correlations further away from the superfluid-supersolid quantum critical point, 
or examining other estimators that would be expected to harbor a complementary anisotropy in such a state.
Since it is also possible that the doped-Mott insulator interpretation simply ceases to be valid at densities this far from 1/3 or 2/3, additional measurements of the $\rho_s$ anisotropy away from 1/2 filling would help shed light on the issue.
The resolution of such paradoxes in a simple model like this could have broad implications for studies of nanoscale density inhomogeneities in other bulk quantum phases, for example in ``stripe'' mechanisms of high-$T_c$ superconductivity in the cuprates.

The 1/3-filled striped Mott insulating state exists on the $V=0$ phase diagram for large $V^{\prime}$, in the presence of a chemical potential that breaks particle-hole symmetry.  In addition to being adjacent to the striped supersolid phase, this phase borders on a uniform superfluid phase that dominates at larger hopping $t$.  This phase boundary offers a unique opportunity to search for an example of exotic non-LGW quantum criticality suggested by a recent phenomenological DVT.\cite{theBurkonian}  However, in this model, the relevant superfluid-Mott phase transition is observed to be strongly first-order.  Thus an example of a highly sought deconfined quantum critical point\cite{deconf1} is not present.  Although it is true that the DVT is a phenomenological theory and does not predict the order of a phase transition for any specific microscopic model, it is perhaps unfortunate that the expected universal properties of the vortex theory are not realized in a simple model such as this.

On the other hand, the result that the relevant superfluid-Mott phase transition is {\it strongly} first order is an important contrast to other phase transitions previously studied as candidates for 
DQCPs.
In particular, simulations of superfluid-VBS phase transitions\cite{kagomeVBS,RingEx} 
which at first appear to be continuous (for example because of the lack of obvious discontinuities in the data) may indeed turn out to be very ``weakly'' first-order upon closer scrutiny of more subtle indicators, such as values of critical exponents or interpretations of finite-size behavior on extremely large lattices.  
In contrast to these previous simulations that involved transitions into the VBS phase, our present candidate transition only requires the presence of a simple density-modulated Mott insulator.  Comparing the numerical signatures between these cases may help develop a more systematic method for identifying weakly-first order phase transitions in quantum simulations, which could contribute to the resolution of fundamental controversies\cite{weak1} surrounding the very existence of DQCPs in these and related systems.



\section{Acknowledgments}
We delight in elucidating discussions with S.~Sachdev, A.~Sandvik, A.~Paramekanti, S.~Isakov, J.~Thomas and D.~Scalapino.  RGM would like to thank the Physics Department at Harvard University for the hospitality extended during a visit, which led to some of the ideas discussed here.  AD acknowledges support from NSERC of Canada and AB acknowledges support from NSF Grant DMR02-33773.  The computer simulations were performed using resources provided by the Harvard Center for Nanoscale Systems, part of the National Nanotechnology Infrastructure Network, and the IBM p690 Power4 machine operated by the NCCS at ORNL.

\end{document}